\documentclass{aastex63}
\pdfoutput=1
\usepackage{graphicx}
\usepackage{mathrsfs,amssymb}
\usepackage{amsmath}
\shorttitle{No evidence for planets orbiting Kapteyn's star}
\shortauthors{}

\begin{document}

\title{A Gaussian Process Regression Reveals No Evidence for Planets
Orbiting Kapteyn's Star}


\author{Anna Bortle}
\author{Hallie Fausey}
\author{Jinbiao Ji}
\author{Sarah Dodson-Robinson}
\author{Victor Ramirez Delgado}
\author{John Gizis}

\affiliation{University of Delaware \\ Department of Physics
and Astronomy \\ 217 Sharp Lab \\ Newark, DE 19716 USA}



\email{anbortle@gmail.com, hfausey@gwmail.gwu.edu, jinbiao@udel.edu, sdr@udel.edu, viclrd@udel.edu, gizis@udel.edu}

\begin{abstract}

  Radial-velocity (RV) planet searches are often polluted by signals caused by gas motion at the star's surface. Stellar activity can mimic or mask changes in the RVs caused by orbiting planets, resulting in false positives or missed detections.
  Here we use Gaussian Process (GP) regression to disentangle the contradictory reports of planets vs.\ rotation artifacts in Kapteyn's star \citep{escude14, robertson15, escude16}.
  To model rotation, we use joint quasi-periodic kernels for the RV and H$\alpha$ signals, requiring that their periods and correlation timescales be the same. We find that the rotation period of Kapteyn's star is 125~days, while the characteristic active-region lifetime is 694~days.  Adding a planet to the RV model produces a best-fit orbital period of 100~years, or 10 times the observing time baseline, indicating that the observed RVs are best explained by star rotation only.
  We also find no significant periodic signals in residual RV data sets constructed by subtracting off realizations of the best-fit rotation model and conclude that both previously reported ``planets'' are artifacts of the star's rotation and activity. 
  Our results highlight the pitfalls of 
  using sinusoids to model quasi-periodic rotation signals.

\end{abstract}



\section{Introduction}\label{sec:intro}

\subsection{Kapteyn's Star}\label{subsec:kapteyn}

\begin{figure} 
    \begin{center}
    \begin{tabular}{c}
    \includegraphics[width=0.8\textwidth]{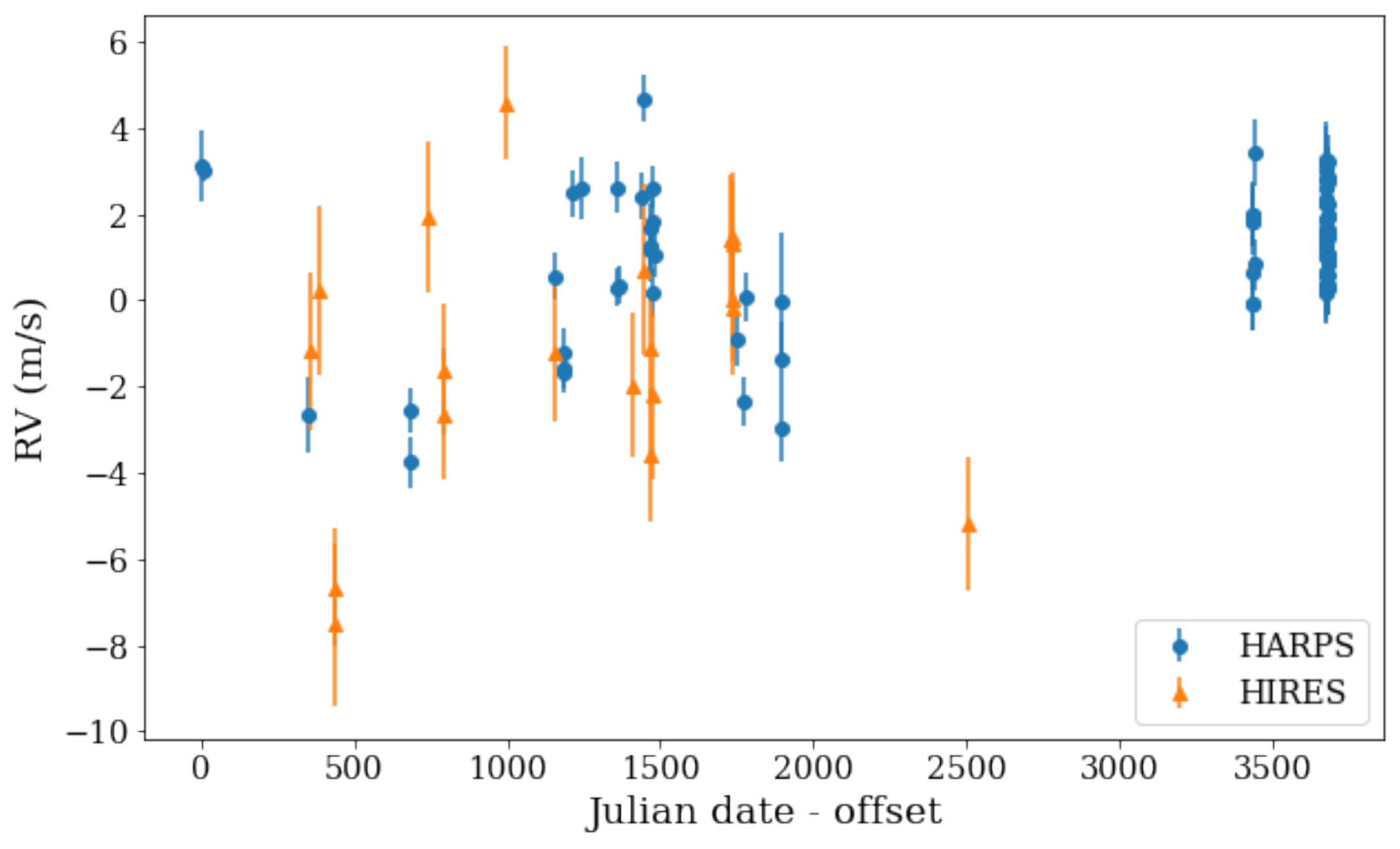} \\
    \includegraphics[width=0.8\textwidth]{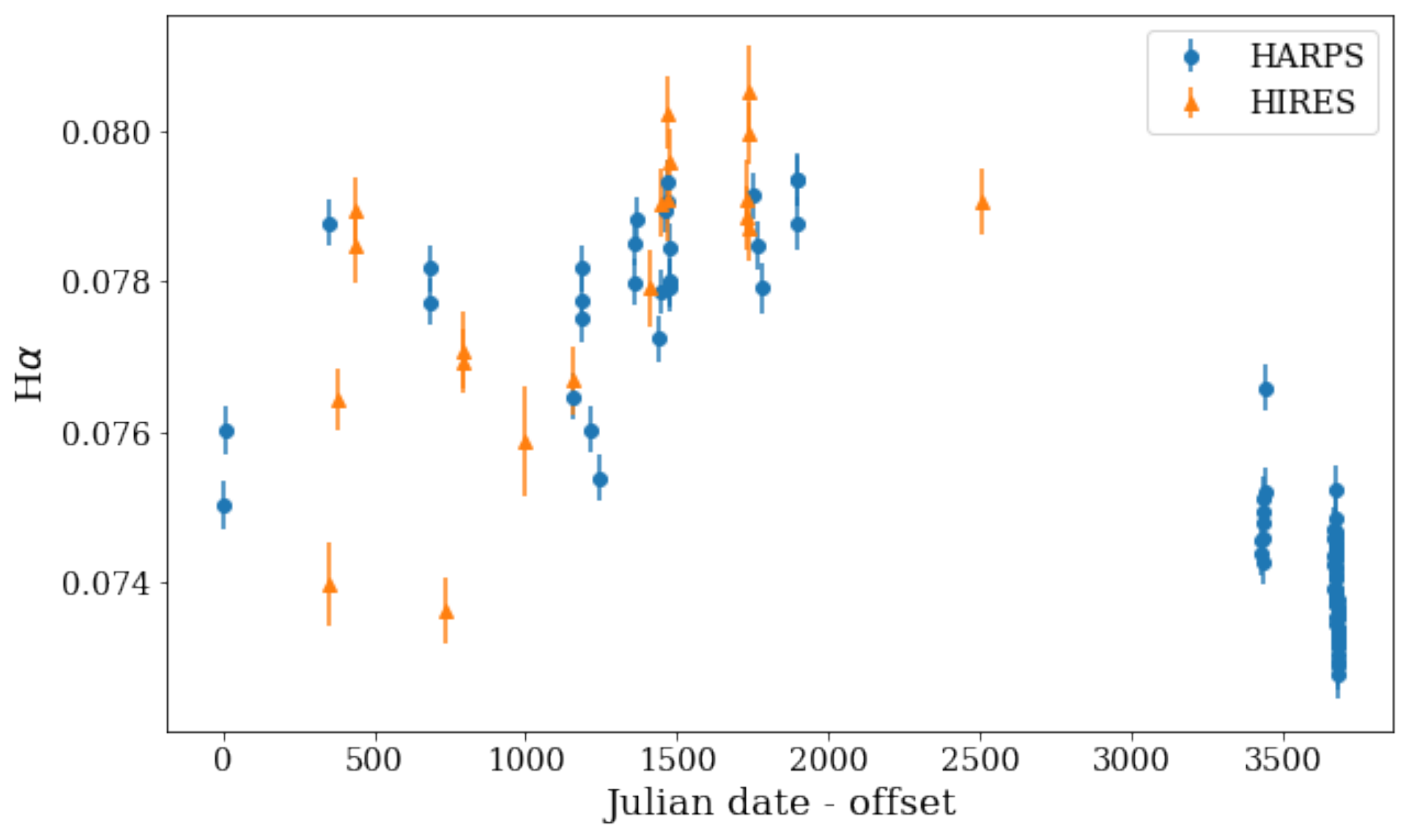}
    \end{tabular}
    \caption{RV (top) and H$\alpha$ (bottom) measurements of Kapteyn's Star from HARPS and HIRES spectrographs.}
    \label{fig:data}
     \end{center}
\end{figure}

In 2014, two exoplanets were reported around Kapteyn's star (GJ 191, HD 33793), a red M1 sub-dwarf \citep{gizis97} located 3.91 parsecs from the Sun (see Table \ref{tab:kapteyn_info}).
The 
reported periods were 48.6 days (planet b) and 121.5 days \citep[planet c;][]{escude14}, with the former falling in the star's habitable zone. The validity of the planets was contested by \citet{robertson15}, who claimed that the star's rotation period is $P_{\rm rot} = 143$ days, and that the 48.6 day period associated with Kapteyn b is an integer fraction (1/3) of $P_{\rm rot}$ and therefore a product of stellar activity. Additionally, they asserted that the 121.5 day period of Kapteyn c is close enough to the star's rotation period to require further observation to determine the planet's legitimacy. Figure \ref{fig:data} shows the RV and H$\alpha$ time series measured by \citet{escude14} and \citet{robertson15}, respectively.

\begin{figure}
    \centering
    \includegraphics[width=0.5\textwidth]{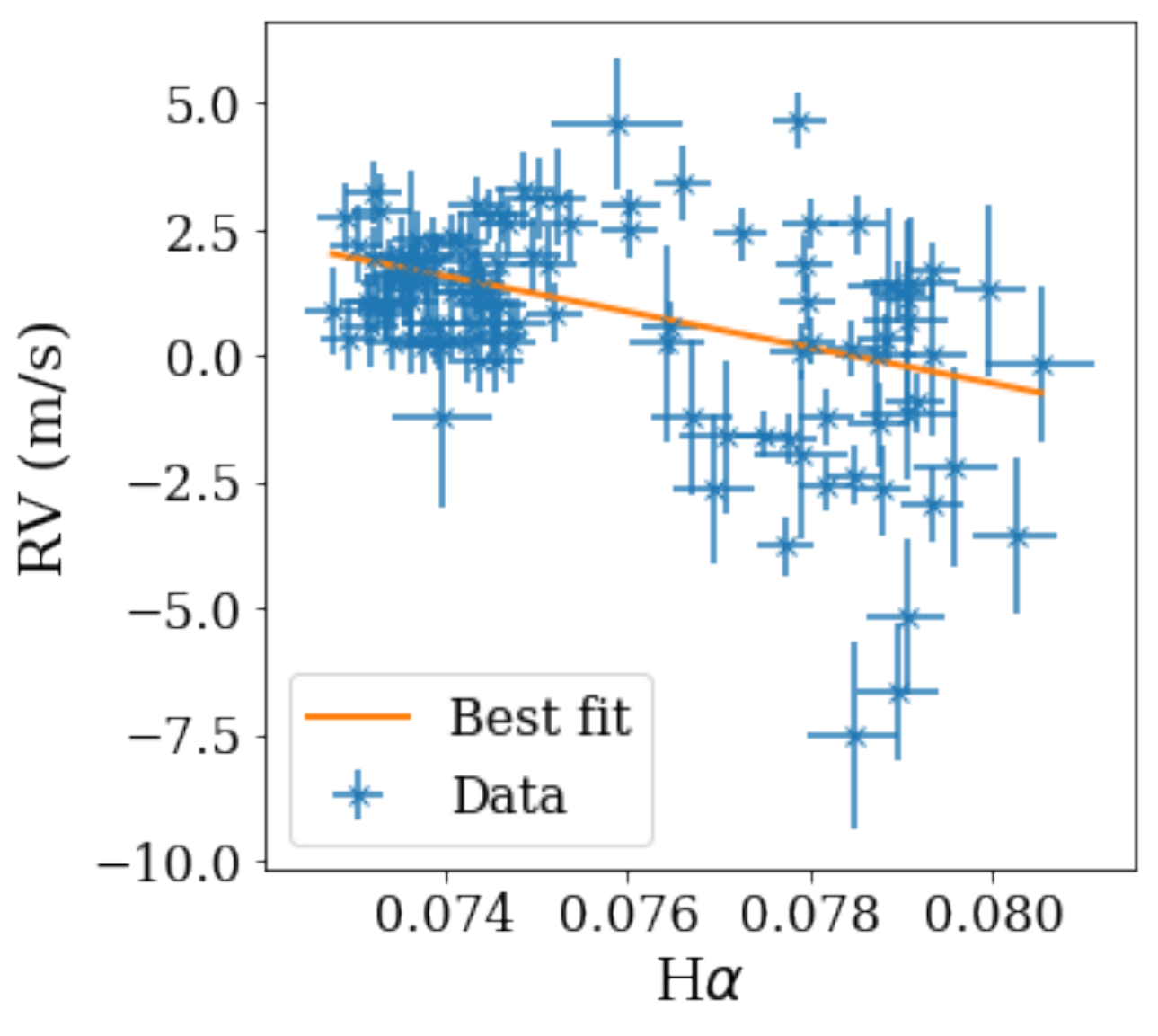}
    \caption{RV as a function of H$\alpha$ (blue crosses with errorbars) and line of best fit found by orthogonal distance regression (orange).}
    \label{fig:linear}
\end{figure}

\citet{escude14} used a log-likelihood periodogram to identify the 48.6-day and 121.5-day signals, then verified that RV could not be described by a straight-line function ${\rm RV} = ax + b$ of either of two activity indicators $x$.
While their chosen activity indicators---V-band photometry and S-index of Ca II H\&K emission---are often helpful with more massive stars, the H$\alpha$ absorption line is more reliable in capturing rotation and stellar activity profiles of dimmer M dwarfs like Kapteyn's star \citep{carolo14, delfosse98, houdebine95, kurster03, stauffer86, suarez15}. Figure \ref{fig:linear} shows RV 
as a function of H$\alpha$ index.
RV and H$\alpha$ appear to be anticorrelated, which means that stellar activity might be manifesting in the RVs. 
The slope of the best-fit line (found by orthogonal distance regression using \texttt{python} package \texttt{scipy.odr} and plotted in orange) differs from zero by more than four standard deviations: $\Delta ({\rm RV}) / \Delta ({\rm H}\alpha) = -355 \pm 72$. 
\citet{robertson15} also noted the statistically significant anticorrelation between H$\alpha$ and RV. However, they fit both quasi-periodic time series with sinusoids, which can result in fictitious periodic signals appearing in the residuals.



Furthermore, Anglada-Escud\'e et al.\ reported finding similar periodicities in the bisector inverse slope (BIS) and S-index as in the RVs, but dismissed their importance due to high false alarm probabilities (FAP) of $\sim$5\%. Because stellar activity is not perfectly sinusoidal but instead quasi-periodic \citep[e.g.][]{nava20}, higher FAPs are expected for stellar activity signals than for planet-induced Doppler shifts. Signals in activity-indicator periodograms with FAP$\sim 5$\% should not be deemed insignificant when they occur on the same periods as RV variations. We therefore choose to fit the star's RV and activity signals using Gaussian Processes with linked quasi-periodic kernels.


\begin{table}
\centering
\begin{tabular}{|l|l|l|ll}
\cline{1-3}
\textbf{Characteristic} & \textbf{Value}  & \textbf{Reference}                                    &  &  \\ \cline{1-3}
Age                     & $\sim$10-12 Gyr & Jofr\'e et al. 2011, Klement 2010, Kotoneva et al. 2005 &  &  \\ \cline{1-3}
Mass                    & $0.281 \pm 0.014 $ M\textsubscript{\(\odot\)}  & S\'egransan et al. 2003                                 &  &  \\ \cline{1-3}
Median H$\alpha$ index    &  $0.07465$               &  Robertson et al. 2015, this work                                                     &  &  \\ \cline{1-3}
Metallicity             & {[}Fe/H{]}$=-0.86$ & Woolf et al. 2005                                     &  &  \\ \cline{1-3}
Radius                  & $0.291 \pm 0.025$ R\textsubscript{\(\odot\)}  & S\'egransan et al. 2003                                 &  &  \\ \cline{1-3}
Temperature             & $3570 \pm 156$ K     & S\'egransan et al. 2003                                 &  &  \\ \cline{1-3}
\textbf{Model Hyperparameter} &  & 
 & & \\ \cline{1-3}
Rotation period: $P$ & $124.71 \substack{+0.19 \\ -0.19}$ days & This work
 & & \\ \cline{1-3}
 Correlation timescale: $\lambda$ & $694.14 \substack{+154.30 \\ -148.53}$ days & This work
  & & \\ \cline{1-3}
  RV kernel amplitude: $A_{RV}$ & $5.95 \substack{+2.62 \\ -1.69}$ m/s & This work
  & & \\ \cline{1-3}
  RV $sin^2$ amplitude: $\Gamma_{RV}$ & $21.50 \substack{+9.62 \\ -6.39}$ m/s & This work
   & & \\ \cline{1-3}
   H$\alpha$ kernel amplitude: $A_{H\alpha}$ & $4.74*10^{-6} \substack{+1.37*10^{-6} \\ -9.69*10^{-7}}$ & This work
  & & \\ \cline{1-3}
  H$\alpha$ $sin^2$ amplitude: $\Gamma_{H\alpha}$ & $144.14 \substack{+104.74 \\ -63.68}$ & This work
   & & \\ \cline{1-3}

\end{tabular}
\caption{Various physical characteristics of Kapteyn's star and its activity cycle.}\label{tab:kapteyn_info}
\end{table}


\subsection{Gaussian Processes}\label{subsec:gaussian}

A Gaussian Process regression involves fitting a physically motivated model to the covariance matrix of a dataset, rather than fitting a function directly to the data. The covariance matrix $k$, which encodes the ``relatedness'' of different pairs of data points, has the form
\begin{equation} \label{eq:covariance}
    k_{ij} = f(t_i - t_j),
\end{equation}
where $f$ is the model function (see \S\ref{sec:GP}) and $t_i - t_j$ is the time lag between observations $i$ and $j$.
One use of the GP is modeling systematic and instrumental error. For example, \citet{gibson12a} utilized the non-parametric nature of the GP to determine transit parameters from transmission spectroscopy of HD 189733 in the presence of systematic noise. Additionally, \citet{czekala15} constructed covariance matrices using GP kernels to correct for artifacts in the residuals of spectroscopic measurements due to inexact models and systematic errors. \citet{luger16} also created an open-source pipeline called EVEREST that employs GP for removing systematic and instrumental noise from K2 light curves.

GP are also effective tools for modeling correlated noise, which is what stellar activity contributes to planet-search data. \citet{barclay15} used GP to model RV noise from granulation in red giants. They found that previous papers claiming that the Kepler-91b signal was a false positive had not fully accounted for stellar activity, and reaffirmed the planet's existence. \citet{angus18} used GP to model the rotation periods of 333 simulated and 275 real Kepler light curves using quasi-periodic kernels to capture both the periodic effects of rotating starspots and the changes in spot coverage over time. They demonstrated that the GP method produced more accurate results than the autocorrelation function and Lomb-Scargle periodogram methods. Furthermore, \citet{damasso17} utilized GP to remove correlated stellar noise from RV curves of Proxima Centauri, and were able to model both the stellar activity and the planetary signal of Proxima b. They confirmed that Proxima's activity signals have a periodic component that is related to the rotation period of the star.

One can model joint datasets (for example, simultaneous RV and H$\alpha$ measurements) using linked GPs, or gain insight into the nature of a single dataset by fitting it with different GP kernel models.
For example, \citet{jones17} used principal component analysis (PCA) to construct high-information activity indicators that include contributions from many activity-sensitive lines, then modeled variations in both RV and the PCA activity indicators using linear combinations of $X(t)$, $\dot{X}(t)$, and $\ddot{X}(t)$, where the unknown function $X$ and its derivatives were described by GPs. Jones et al.\ were extending the work of \citet{rajpaul15}; we will compare our methods with the pioneering studies of Rajpaul et al.\ and \citet{haywood14} in more detail in the conclusion. Additionally, \citet{pope16} modeled stellar variability in K2 planet-search datasets using either a quasi-periodic or a slowly varying exponential GP kernel, depending on whether the Lomb-Scargle periodogram of each light curve showed significant periodicity. Their approach was particularly effective for validating planet candidates orbiting young variable stars. GP have also been adopted to revisit the occurrence rate density of Earth-like planets, with the ability to account for observational noise and implement fewer but more secure assumptions \citep{foreman-mackey14}.

On the planet discovery side, \citet{cloutier17} determined the orbital parameters of the known super-earth K2-18b with GP and uncovered evidence for another potential planet around the host star. Moreover, \citet{czekala17} used GP for disentangling spectral lines from the two stars of eclipsing binary system LP661-13, and were able to successfully separate the signals while accounting for the changing redshifts at various points in the stars' orbits. The signal morphologies were consistent with the types of stars in the binary system.
Other examples of GP validation of planet signals in RV data include the detections of new planets K2-131b \citep{dai17}, K2-291b \citep{kosiarek19a}, a super-Earth around M-dwarf G1 686 \citep{affer19}, and a sub-Neptune around HD 175607 \citep{mortier16}. Additional stellar, planet, and transit characterizations using GP are \citet{gibson12b, gibson13}, \citet{haywood14, haywood18}, \citet{damasso18}, \citet{farr18}, \citet{kosiarek19b}, and \citet{suarezmascareno20}.

In an attempt to resolve the debate on the exact rotation period of Kapteyn's star and the extent to which its rotation and activity materialize in the RV data, we adopt GP to model the star's rotation. We use joint quasi-periodic kernels for the RV and H$\alpha$ signals. By modeling the two together, we can disentangle activity and rotation from any exoplanet-related RV signals. 

\section{Gaussian process model}\label{sec:GP}

\begin{figure} 
    \begin{center}
    \includegraphics[width = \textwidth]{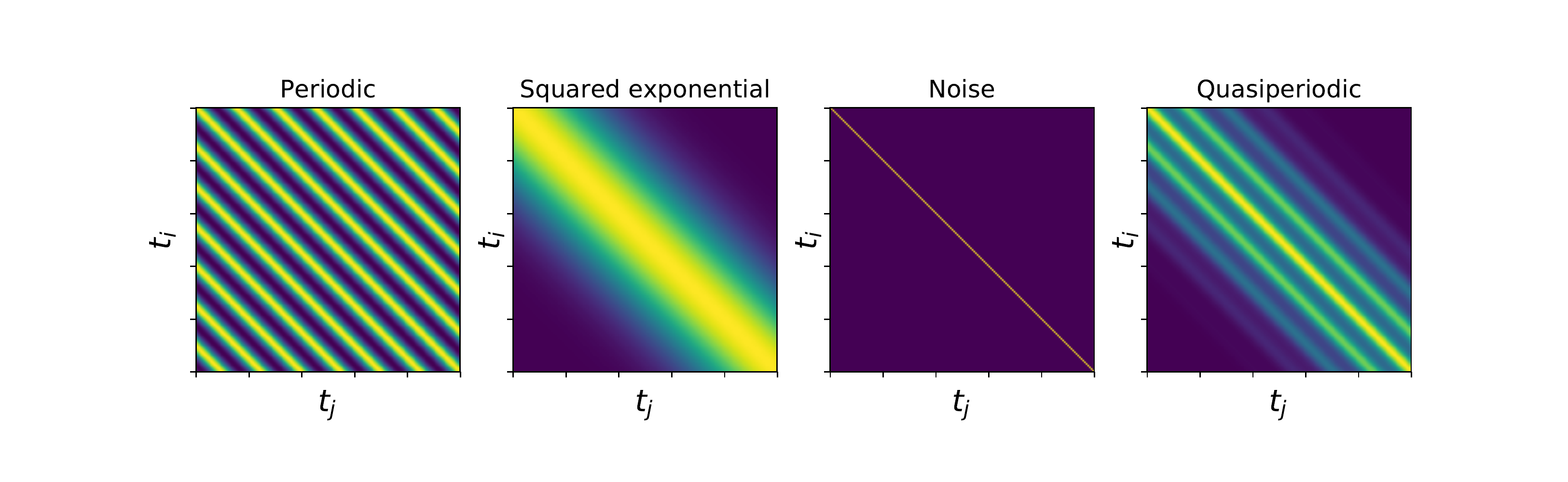}
    \caption{Visualizations of each component of the quasi-periodic kernel and the final kernel.}
    \label{fig:kernel_visual}
     \end{center}
\end{figure}

\subsection{The GP kernel}
\label{subsec:kernel}

A GP is defined by its mean and covariance functions.
It is common practice to assume that the mean function is zero everywhere, so choosing an appropriate covariance model, or kernel for the GP, is essential.
As shown in Figure \ref{fig:kernel_visual}, our model kernel includes a periodic component to capture the star's rotation and a squared exponential component to account for the changing configuration of active regions. The squared exponential kernel encodes a weaker correlation between data points that are far from each other in the time series (high $|t_i - t_j|$) and a stronger correlation between those that are closer (low $|t_i - t_j|$). Multiplication of the two GPs produces a quasi-periodic kernel. Finally, white noise terms were added in quadrature to the diagonal entries of the kernel to account for uncertainties in the observed data. 

The equation describing the quasi-periodic kernel is 
\begin{equation} \label{eq:quasi_kernel}
  k_{i,j} = A\ \mathrm{exp}\left[-\frac{(t_i - t_j)^2}{\lambda^2} - \Gamma \mathrm{sin}^2\left(\frac{\pi(t_i -t_j)}{P}\right)\right] + \sigma_i^2\delta_{ij},
  \end{equation}
where $A$ is the amplitude; $\lambda$ is the correlation timescale, which is likely set by the characteristic lifetime of the star's active regions \citep{angus18} but which may also reflect differential rotation; $\Gamma = \frac{1}{2\omega^2}$, where $\omega$ is the smoothness of the periodic component; $P$ is the rotation period; and $\sigma$ is the white noise. We use the term ``characteristic lifetime'' to describe $\lambda$ because large spots persist longer than small spots \citep{bradshaw14}, so it is too simplistic to imagine that every active region has lifetime $\lambda$. Rather, $\lambda$ is an ensemble average over all active regions.
The data we use are the RVs and H$\alpha$ measurements quoted by \citet{escude14} and \citet[Figure \ref{fig:data}]{robertson15}. 
The dataset includes 92 HARPS spectra \citep{pepe00} and 20 HIRES spectra \citep{vogt94}, spanning a course of 10.1 years.

When creating the model, we also considered the possibility that a long-term stellar activity trend such as the sun's 11-year cycle might manifest in the time interval covered by the data set.
Thus we had two possible models for the H$\alpha$ variability---one with the quasi-periodic kernel given by Equation \ref{eq:quasi_kernel} to describe RV and H$\alpha$ due to rotation only, and one with Equation \ref{eq:quasi_kernel} describing RV but an additional squared exponential kernel applied to H$\alpha$ to model long-term signal drift:
 \begin{equation} \label{eq:ha_lt_kernel}
 k_{i,j} = A\ \mathrm{exp}\left[-\frac{(t_i - t_j)^2}{\lambda^2} - \Gamma \mathrm{sin}^2\left(\frac{\pi(t_i -t_j)}{P}\right)\right] + B\ \mathrm{exp}\left[-\frac{(t_i - t_j)^2}{\gamma^2}\right] + \sigma_i^2\delta_{ij}.
 \end{equation}
 In Equation \ref{eq:ha_lt_kernel}, $B$ is the squared exponential amplitude for the long-term activity cycle kernel, $\gamma$ is the correlation timescale for the long-term H$\alpha$ activity kernel, and all other variables are as defined above in Equation \ref{eq:quasi_kernel}. The extra H$\alpha$ signal drift described by Equation \ref{eq:ha_lt_kernel} is general enough to cover any stellar activity that appears in H$\alpha$, but is not obvious in RV. The motivation to add the long-term drift to H$\alpha$ only, not RV, comes from examining the periodograms of both time series. In the 
 activity-indicator periodograms presented by \citet{robertson15} (their Figure 1), the most powerful signals in H$\alpha$, Na~D, FWHM, and S-index are at $P > 3000$~days. In contrast, the RV periodogram presented by \citet{escude14} has no long-period signals that approach the power of the 121-day peak identified as planet c.
 
 In both potential rotation / activity models, the periods $P$ and correlation timescales $\lambda$ were constrained to be identical for the RV and H$\alpha$ data sets. Since H$\alpha$ is a good indicator of stellar activity, this requirement ensured that the GP was modeling the rotational and activity signals of Kapteyn's star as they manifest in the velocities, and not identifying other unrelated signals as stellar activity. When RVs trace stellar activity or rotation, it might seem reasonable to assume that the activity-indicator and RV time series have the same frequency domain structure; in that case, both GP covariance kernels should have the same value of $\Gamma$ \citep{haywood14}. However, we found that the GP fit would not converge when we constrained the RV and H$\alpha$ to kernels to have identical $\Gamma$. We discuss the physical implications of RV and H$\alpha$ having different frequency-domain structures, but the same dominant period, in Section \ref{sec:disc}.
 
Finally, we designed one more numerical experiment to evaluate the \citet{robertson15} speculation that the observed RVs could have contributions from both rotation and a planet candidate. Here we fit the RVs with a combination of the GP rotation model and a Keplerian orbit (no long-term drift in H$\alpha$ is included). In the rotation+planet experiment, the GP kernels are unchanged from the pure rotation model---they are still defined as in Equation \ref{eq:quasi_kernel}, with identical periods and correlation timescales for RV and H$\alpha$. The best-fit planet orbital parameters are found by fitting the GP not to the observed RVs, but rather to the residual RVs obtained by subtracting off a Keplerian orbit, as described in \S \ref{subsec:modelfitting}.


\begin{figure} 
    \begin{center}
    \includegraphics[scale = 0.8]{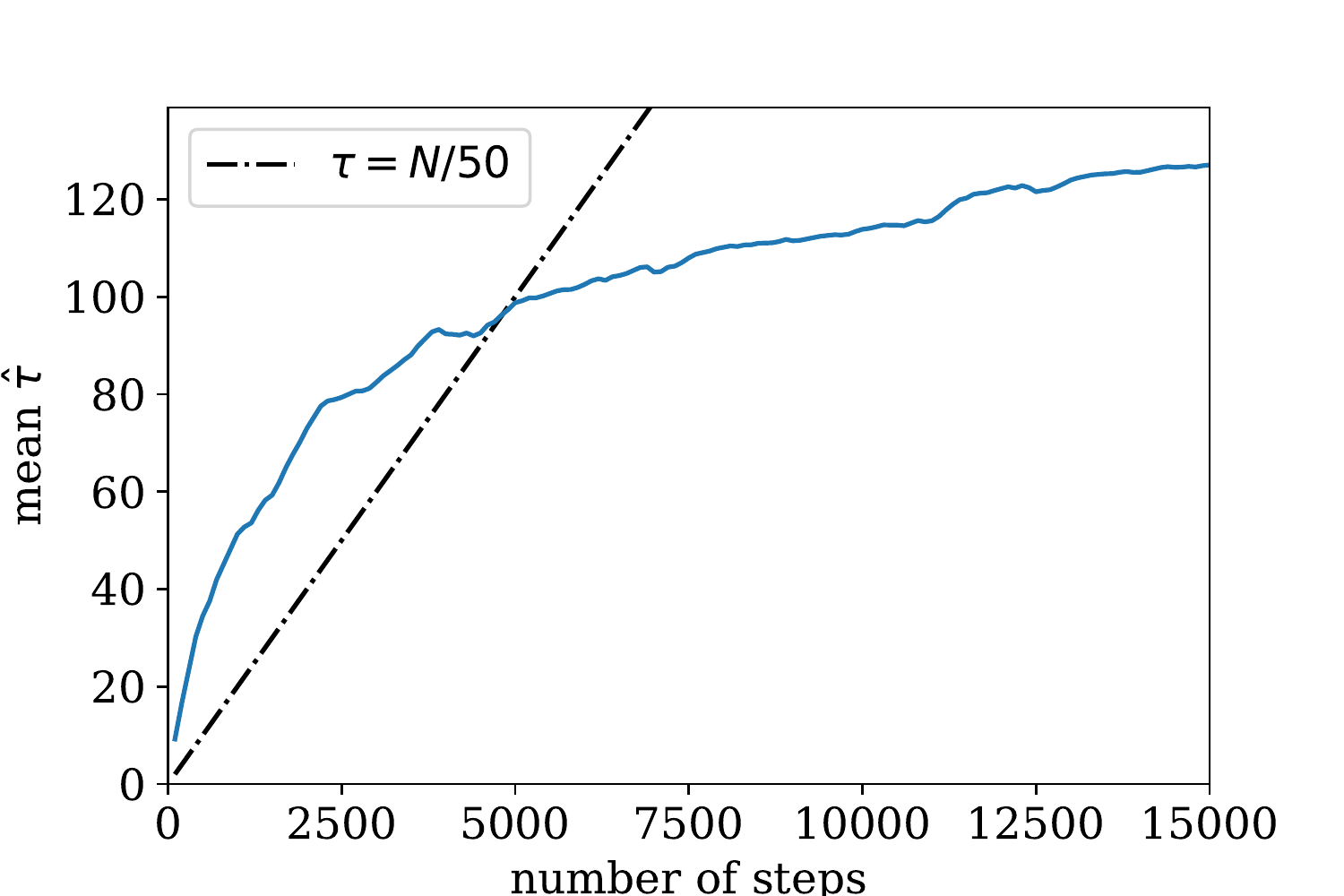}
    \caption{Estimated autocorrelation time (blue) with the $\tau = N/50$ line (black) for finding convergence of the GP. Mean $\hat{\tau}$ is computed by averaging over the integrated autocorrelation time estimator (Equation \ref{eq:tauest}) for all hyperparameters.}
    \label{fig:convergence}
    \end{center}
\end{figure}

\subsection{GP fitting and MCMC simulations}
\label{subsec:modelfitting}

The \texttt{george}\footnote{\texttt{george} documentation can be found at \url{https://george.readthedocs.io/en/latest/}} \citep{ambikasaran15, george} and \texttt{emcee}\footnote{\texttt{emcee} documentation can be found at \url{https://emcee.readthedocs.io/en/latest/}} \citep{emcee} packages were used to implement the GP and sample the posterior probability distribution functions for the two model kernels. Informative initial guesses for the kernels' hyperparameters were made from the periods and RV/H$\alpha$ amplitudes reported in \citet{escude14} and \citet{robertson15}. The hyperparameters were set with uniform prior windows: $-2 \leq \ln{P} \leq 10$, $0 \leq \ln{\lambda} \leq 18$, $-4 \leq \ln{A_{RV}} \leq 6$, $1.0 \times 10^{-5} \leq \ln{\Gamma} \leq 100$, $-20 \leq \ln{A_{H\alpha}} \leq 0$, $-20 \leq B \leq 0$, and $0 \leq \ln{\gamma} \leq 23$. 

\texttt{George}'s built-in log-likelihood capabilities were paired with \texttt{scipy.optimize.minimize}\footnote{\texttt{scipy.optimize.minimize} documentation can be found at \url{https://docs.scipy.org/doc/scipy/reference/generated/scipy.optimize.minimize.html}} to find the maximum log-likelihood values for the GP hyperparameters. For the joint RV and H$\alpha$ system, the log-likelihood is a sum of the separate RV and H$\alpha$ log-likelihoods from their respective GP kernels: 
\begin{equation}
    L_{\rm model} = L_{\rm RV} + L_{H\alpha}.
    \label{eq:combinedlikelihood}
\end{equation}
From \citet{rasmussen05}, the log-likelihood of the GP model described by kernel $K$ is
\begin{equation} \label{eq:loglikelihood}
    L = 
    -\frac{1}{2}\mathbf{y}^T(\mathbf{K}+\sigma^2_{i}\mathbf{I})^{-1}\mathbf{y} - \frac{1}{2}\ln |\mathbf{K}+\sigma^2_{i}\mathbf{I}| - \frac{n}{2}\ln 2\pi,
\end{equation}
where 
$\mathbf{y}$ is the vector containing the observed data,
$\mathbf{K}$ is the covariance matrix/kernel with determinant $|\mathbf{K}|$, $\sigma_i$ are the error bars on each measurement $i$, $n$ is the number of data points, $\mathbf{I}$ is the identity matrix, and $^T$ denotes transpose. When the Keplerian orbit is included in the RV model, we have
\begin{equation}
L_{\rm planet} = -\frac{1}{2} \mathbf{r}^T \left( \mathbf{K} + \sigma_i^2 \mathbf{I} \right)^{-1} \mathbf{r} - \frac{1}{2} \ln | \mathbf{K} + \sigma_i^2 \mathbf{I} | - \frac{n}{2} \ln (2 \pi)
\label{eq:loglike_planet}
\end{equation}
\citep[e.g.][]{cabot21}. In Equation \ref{eq:loglike_planet}, $\mathbf{r}$ is the residual after fitting a Keplerian orbit:
\begin{equation}
    \mathbf{r} = \mathbf{RV} - m(t),
    \label{eq:Keplerian_resids}
\end{equation}
where $\mathbf{RV}$ is the vector of radial velocities and $m(t)$ is the Keplerian model.
The hyperparameters of the maximum log-likelihood model were used to initialize the \texttt{emcee} Markov-chain Monte Carlo (MCMC) sampler. 



\subsection{MCMC convergence}
\label{subsec:convergence}

To ensure enough steps were taken in the MCMC sampler chain, we used \texttt{emcee} 
to investigate the number of autocorrelation times covered by different chain lengths. 
$\tau_N$, the {\it integrated} autocorrelation time, is the number of steps required for the chain to ``forget'' where it started, and helps determine the statistical error in the measurements taken from the chain \citep{sokal97}.
As described in the \texttt{emcee} package documentation \citep{emcee,sokal97}, 
\begin{equation} \label{eq:tau}
    \tau_N = \sum_{\tau = - \infty}^{\infty}\rho_N(\tau)
\end{equation}
where $\rho_N(\tau)$ is the normalized autocorrelation function for a chain of length $N$ as a function of shift $\tau$. $\rho_N(\tau)$ can be estimated as
$$\hat{\rho}_N(\tau) = \frac{\hat{c}_N(\tau)}{\hat{c}_N(0)}$$
where $\hat{c}_N(\tau)$ is the sampler chain's autocorrelation.
The \texttt{emcee} package then estimates $\tau_N$ to be
\begin{equation}
\hat{\tau}_f(M) = 1 + 2\sum_{\tau=1}^M\hat{\rho}_N(\tau)
\label{eq:tauest}
\end{equation}
with $M\ll N$. \texttt{emcee} uses $M$ instead of $N$ to estimate $\tau_N$ to reduce variance that can occur as $\tau$ approaches $N$. Converged chains have $M \geq C\tau_N(M)$ for some constant $C$; 
\citet{emcee} suggest $C=50$ as the MCMC convergence threshold. When the estimated integrated autocorrelation time crosses $\tau_N=N/50$ as shown in Figure \ref{fig:convergence}, the posterior probability distributions for all hyperparameters are adequately sampled. 

The highest $\tau_N$ value among the hyperparameters is that of the period $P$, with $\tau_N = 135$ steps. Therefore, to assure that all parameters converge, the chain lengths must be at least $50 \times \tau_{N,max} = 50 \times 135 = 6750$ steps. We used 50 walkers with 8000 steps, following an initial 500 steps of burn-in, so our Markov chains are long enough to ensure all hyperparameters have surpassed the required $50\tau_N$ chain length and the GP has converged. More information on autocorrelation analysis and convergence can be found in the \texttt{emcee} documentation. \footnote{\texttt{emcee} documentation on autocorrelation analysis and convergence can be found at \url{https://emcee.readthedocs.io/en/latest/tutorials/autocorr/\#}}

\section{Results: Rotation, best-fit Keplerian, and search for residual periodicity}\label{sec:results}

\subsection{Choosing the best-fit stellar activity model}
\label{subsec:bestactivity}

The best model of stellar activity (quasi-periodic only vs.\ quasi-periodic plus long-term drift in H$\alpha$) was decided by examining the 
Bayesian Information Criterion (BIC), likelihood ratio (LR), and average reduced $\chi^2$ from each model.
We calculated the BIC for all GP realizations in the \texttt{emcee} sampler chains: 
\begin{equation}\label{eq:BIC}
    {\rm BIC} = \ln({n})k - 2\ln({L}),
\end{equation}
where $k$ is the number of hyperparameters (6 for the quasi-periodic model, 8 for the quasiperiodic+long-term drift model).
By examining a histogram of BIC values from each  MCMC sampler chain, 
we found that 
the more complex model (quasi-periodic in RV, quasi-periodic + long-term  trend in H$\alpha$) often gave lower BIC values than the simpler model (quasi-periodic in RV \& H$\alpha$).
However, the distribution of BIC values for the complex model was wide compared to the distribution for the simpler model---i.e.\
the more complex model had many realizations that fit the data extremely poorly, while the simpler model did not---making the typical BIC value for the complex model comparable to that of the simpler model.

Likelihood ratios were also calculated for the pair of sampler chains, defined by
\begin{equation}\label{eq:LR}
    {\rm LR} = 2(L_1 - L_2),
\end{equation}
where $L_1$ is the log-likelihood of the simpler model and $L_2$ is the log-likelihood of the more complex model. We used LR to 
determine whether the complex model provided enough improvement in the fit to warrant the addition of its extra hyperparameters. The LR values for the each of samples and the difference in number of degrees of freedom (DOF) between the two models were fed to \texttt{scipy}'s survival function (SF), \texttt{scipy.stats.chi2.sf}\footnote{\texttt{scipy.stats.chi2.sf} documentation can be found at \url{https://docs.scipy.org/doc/scipy/reference/generated/scipy.stats.chi2.html}}, to determine which model was a better fit based on its log-likelihood value and DOF. 
\texttt{Scipy}'s SF is defined as $1-$CDF, where CDF is the cumulative distribution function. When given LR and DOF values, the SF returns the probability that the LR will take on a value greater than $0$ which, by definition of LR in Equation \ref{eq:LR}, indicates the probability that the log-likelihood of the simpler model will be greater than that of the model with the additional exponential trend in H$\alpha$.
We found that LR+SF results tended to favor the more complex model.

BIC and LR results both tentatively favored the model with a long-term drift in the H$\alpha$ signal, but also revealed the fragility of the model. With just a small change in the initial hyperparameter guesses, the models' BIC and LR values could worsen such that the simpler model would be preferred. 
The simpler model had LR and BIC distributions that were robust to changes in the initial hyperparameter guesses. Average reduced $\chi^2$ values were calculated using 1000 random samples of each of the models. The quasi-periodic, rotation-only model had an average reduced $\chi^2$ value of 1.820 and the model with the extra long-term drift in H$\alpha$ had an average reduced $\chi^2$ value of 1.884.

The hyperparameter posterior distributions provide the final piece of information that drives us to choose Model 1 (quasiperiodic) over Model 2 (quasiperiodic + long-term drift in H$\alpha$). The posteriors of Model 2 show a high fraction of ``nonsense'' correlation times, where $\lambda \ll P$. These nonsense $\lambda$ values are associated with values of $\gamma$ in the range $\sim 150$--700~days, which are more appropriate for the active-region lifetime than the length of the activity cycle. While the $\gamma$ posterior distribution from Model 2 has a median of 3041 days---a reasonable timescale for an M-dwarf dynamo \citep{robertson13}---the presence of a large number of MCMC samples in which $\lambda \ll P$
means that too often, the quasiperiodic part of the kernel in Equation \ref{eq:ha_lt_kernel} just generates random noise. For Model 1 (Equation \ref{eq:quasi_kernel}), the posterior distributions are always well behaved and the most probable hyperparameter values are always realistic.

Due to the sensitivity of the combined rotation/long-term drift model to the initial guesses, the consistent performance of the rotation-only model in the BIC and LR analysis, and the average reduced $\chi^2$ values, we chose the simpler, exclusively quasi-periodic, rotation-only model to describe the stellar activity. It's important to note that our rotation-period measurement is informed by both H$\alpha$ and RV---we don't have a situation where the best-fit model is primarily constrained by just one of the two time series. When fitting quasiperiodic GPs to RV and H$\alpha$ separately, we found almost identical rotation periods.

\subsection{No evidence of a planet orbiting Kapteyn's star}
\label{subsec:noplanet}

\begin{figure}
    \centering
    \begin{tabular}{cc}
    \includegraphics[width=0.49\textwidth]{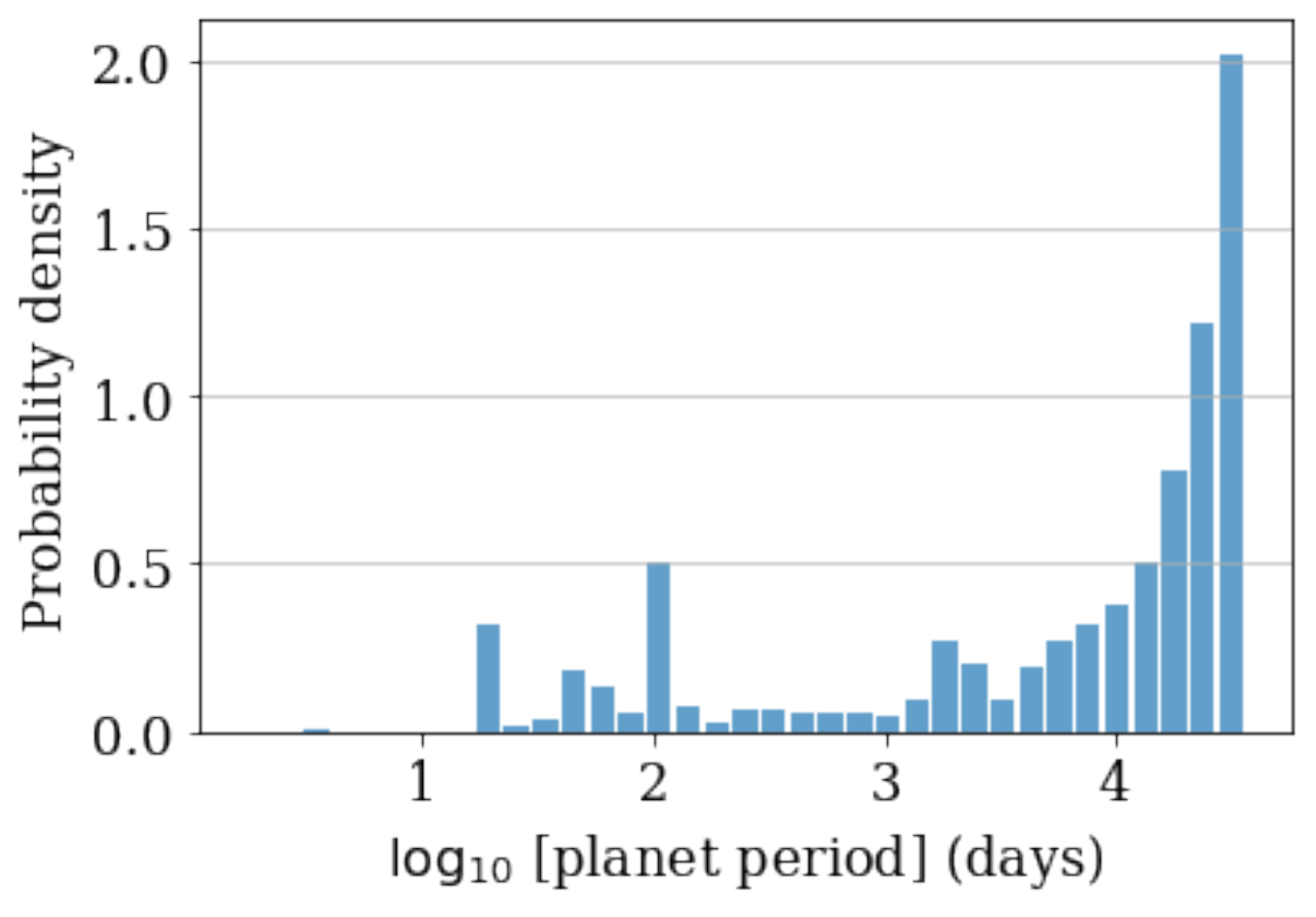} & 
    \includegraphics[width=0.49\textwidth]{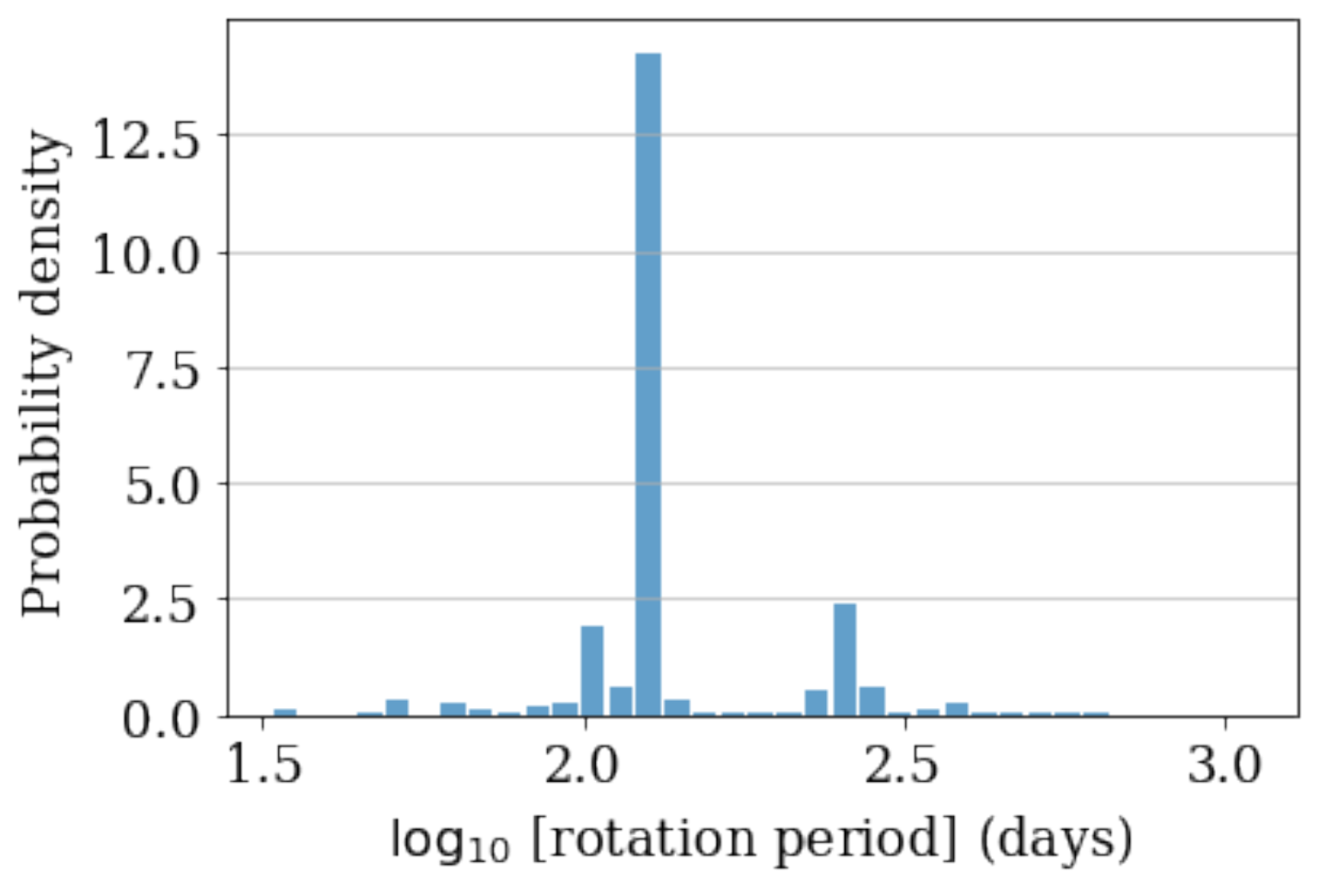} \\
    \end{tabular}
    \caption{{\bf Left:} Posterior distribution of planet orbital periods reconstructed from MCMC exploration of the combined rotation+planet model. {\bf Right:} Posterior distribution of rotation periods from the rotation+planet model.}
    \label{fig:planet_periods}
\end{figure}

When testing the combined rotation+planet model, we almost always got a nonsense answer for the best-fit orbital period. 
In our MCMC simulations, planet period was given a uniform prior with 1~day$ \leq P_{\rm planet} \leq 100$~years. The left panel of Figure \ref{fig:planet_periods} shows the posterior distribution of orbital periods from our MCMC simulations. The most probable orbit takes 100~years, the longest allowed by the prior. Since the observations cover only 10~years, the best-fit planet+rotation model is one in which the planet induces negligible RV variations---i.e., no planet.
When we set prior boundaries that enforced shorter orbital periods, both \texttt{scipy.optimize} and \texttt{emcee} failed to converge on a set of best-fit GP hyperparameters (where \texttt{emcee} convergence was evaluated as described in \S \ref{subsec:convergence}).

The right panel of Figure \ref{fig:planet_periods} shows the posterior distribution of star rotation periods in the combined planet+rotation model. We recover the same rotation period---125 days---as we do when fitting Model 1, rotation only (see discussion of the best-fit hyperparameters in \S \ref{subsec:posteriors}, below). The best-fit rotation+planet model is behaving as if it were a model of rotation only.

Our numerical experiments show that there is no clear evidence for a planet orbiting Kapteyn's star. If planets do exist, they are not revealed by the RV measurements of \citet{escude14}: the RV signal can be fully explained by star rotation. Planet discovery awaits the collection of more data at higher precision.

\subsection{Posteriors and best-fit hyperparameters of the rotation model}
\label{subsec:posteriors}

\begin{figure} 
    \begin{center}
    \includegraphics[width = 0.9\textwidth]{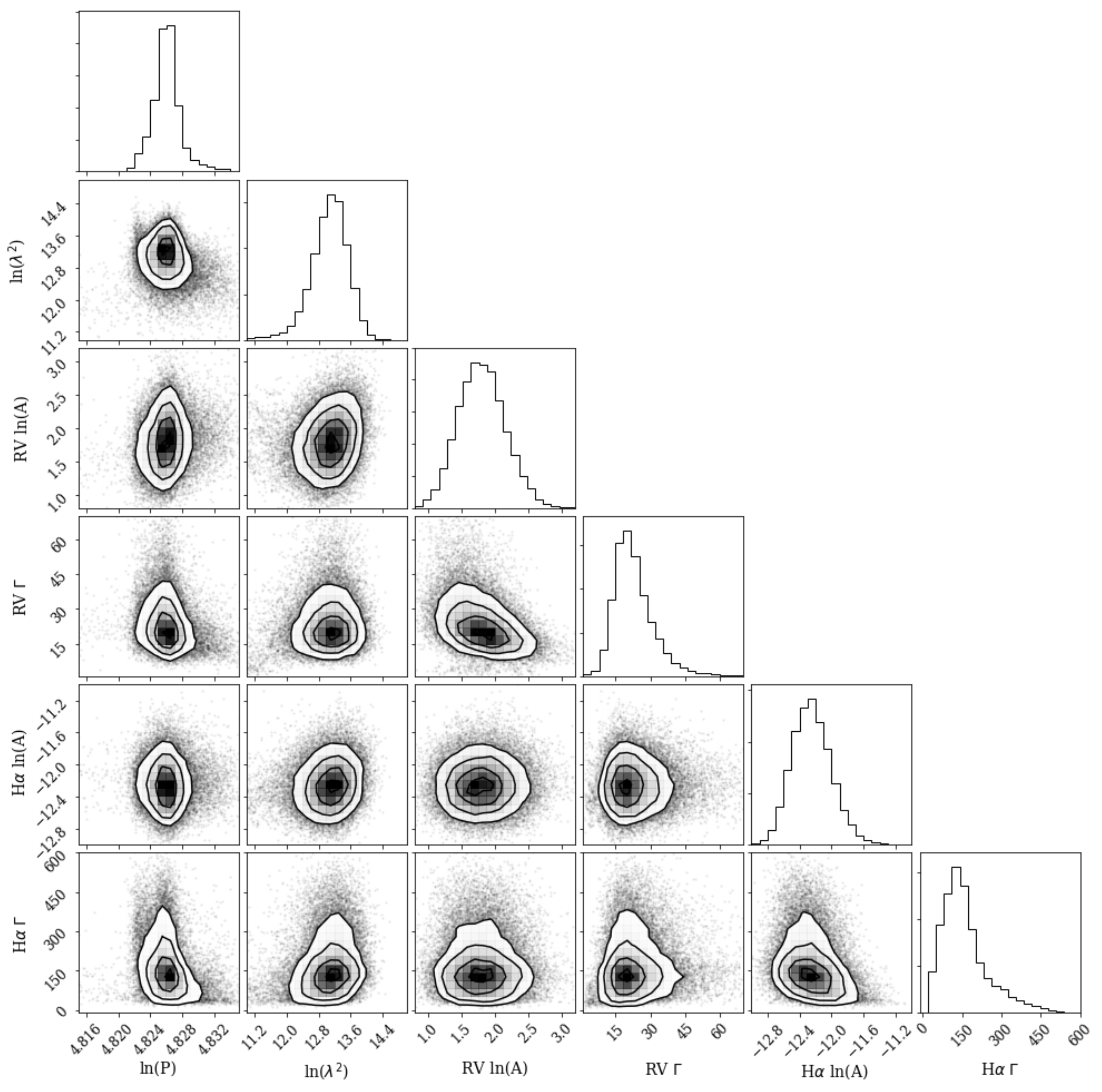}
    \caption{Hyperparameters and posterior distributions from the RV-H$\alpha$ joint quasi-periodic model, showing the distribution of 18,000 samples. Plot generated with \texttt{corner.py}.\footnote{\texttt{https://github.com/dfm/corner.py/blob/main/docs/index.rst}}}
    \label{fig:qsipost}
     \end{center}
\end{figure}

The posterior distributions of the hyperparameters from Model 1, shown in Figure \ref{fig:qsipost}, were examined to see how well the GP found converged values. 
The best-fit period, from the median of the marginalized posterior distribution, is $\ln(P) = 4.826$, corresponding to a period $P = 124.71 \pm 0.19$ days (Table \ref{tab:kapteyn_info}). Our analysis indicates that Kapteyn's star rotates similarly to Barnard's star, a 10 Myr-old M dwarf with $P = 145$~days \citep{toledopadron19}.

\begin{figure}
    \centering
    \includegraphics[width=0.99\textwidth]{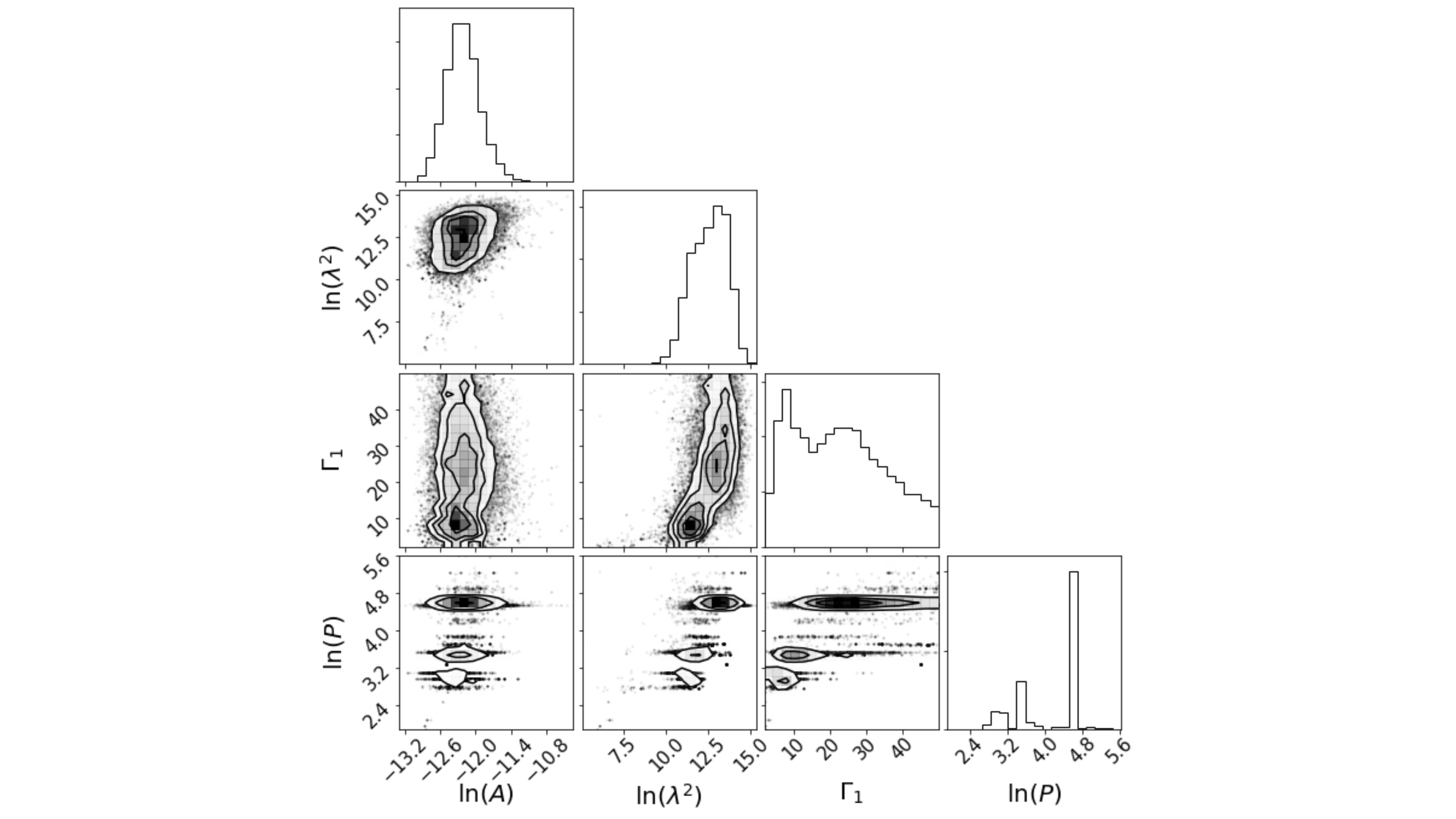}
    \caption{Hyperparameter covariances and marginalized posterior distributions from fitting the quasiperiodic GP (Equation \ref{eq:quasi_kernel}) to the H$\alpha$ time series alone. The trimodality in the $\ln(P)$ posterior disappears when the RV and H$\alpha$ time series are fitted jointly.}
    \label{fig:halpha_corner}
\end{figure}

The period appears to be particularly well constrained, with uncertainties of only 0.15\%. The fact that the period uncertainties are so small is surprising, given the sparsity of the dataset. However, the rotation period we found is quite similar to the frequency of maximum power in the Lomb-Scargle periodogram, which is 121.5 days \citep{escude14, robertson15}. Furthermore, we have tried a variety of initial guesses for the GP optimization, which in turn provides the starting point for the MCMC simulation, and \texttt{emcee} always find a very strong maximum in the log-likelihood at $P = 124.71$~days. When we fit the H$\alpha$ and RV time series separately, the posterior distribution of the period of H$\alpha$ oscillations is trimodal, retaining the peak at $P = 125$~days but with additional small peaks centered at 16.4 and 29.9 days (Figure \ref{fig:halpha_corner}). The trimodality disappears in the joint H$\alpha$-RV GP model, indicating that jointly fitting both observables gives tighter period constraints than modeling either H$\alpha$ or RV alone.

The characteristic active-region lifetime is $\lambda = 694.14$ days. The uncertainties on $\lambda$ are +22\%/-21\% (Table \ref{tab:kapteyn_info}), similar to the correlation timescale uncertainties of +23\%/-17\% found by \citet{damasso17} in their modeling of the rotation of Proxima Centauri. Our ratio of characteristic active-region lifetime to period, $\lambda / P = 5.6$, is consistent with previous results showing that M dwarfs have longer-lived spots than sunlike stars \citep{mathur14}, which typically have $\lambda / P \sim 2$. For example, \citet{robertson15b} found that minor spots on on the M2 dwarf GJ~176 persist for 2--3 rotation periods, while a major spot can last for many years.

Additionally, for the RV kernel we find $\ln$(A$_{RV}) = 1.678$ translating to a variation amplitude of $5.95$, and $\Gamma_{RV} = 19.5$. The best-fit H$\alpha$ kernel has $\ln$(A$_{H\alpha}) = -12.26$, giving a variation amplitude of $4.74 \times 10^{-6}$, and $\Gamma_{H\alpha} = 144.14$. We report these values with errors representing a 68\% credible region (see Table \ref{tab:kapteyn_info}). The GP gives similar results for a variety of initial guesses. This affirms that the GP is convergent toward a specific set of hyperparameters.

\subsection{Model time series and residuals}
\label{subsec:resids}

\begin{figure} 
    \begin{center}
    \begin{tabular}{c}
    \includegraphics[scale = 0.60]{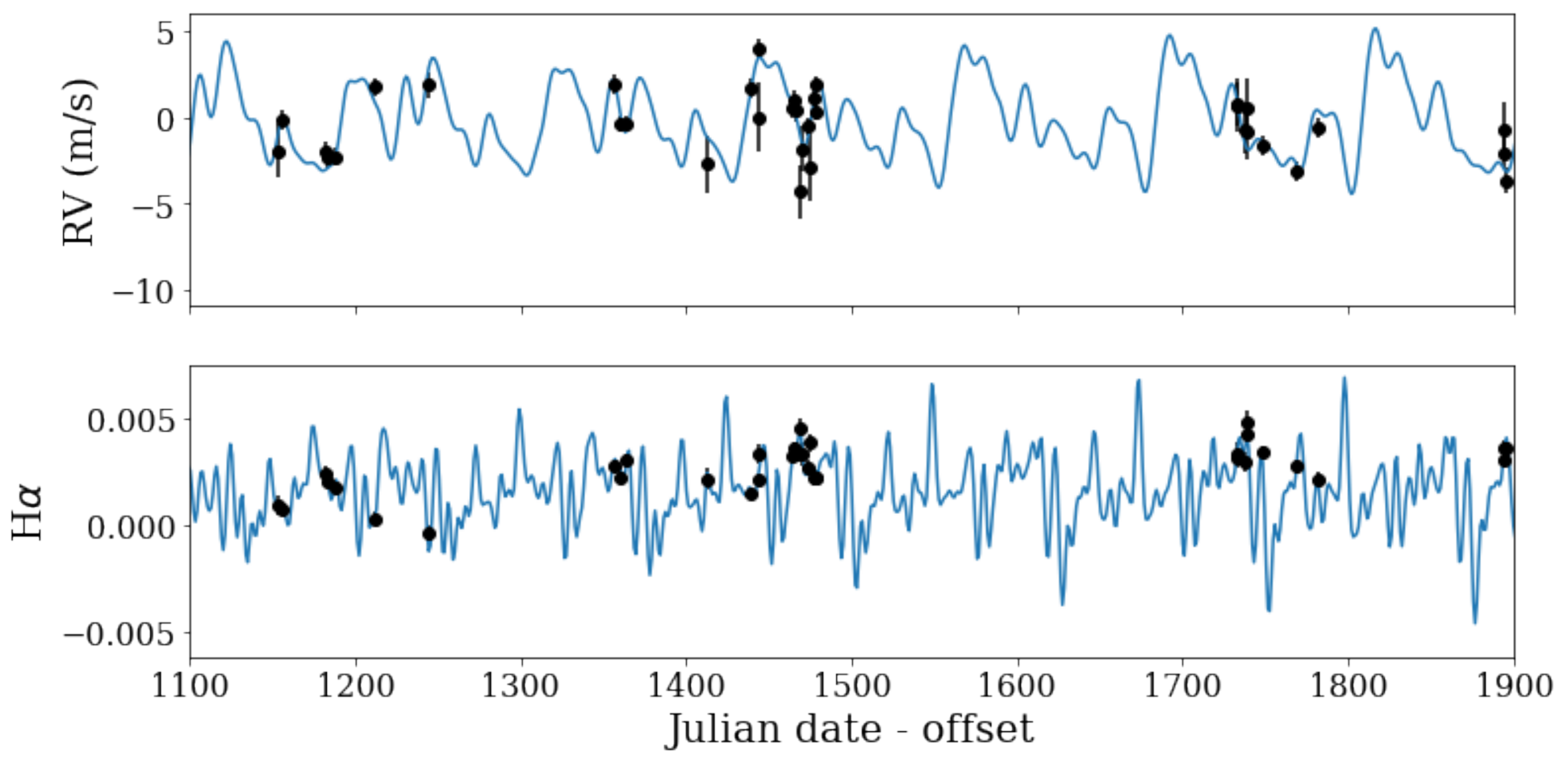} \\
    \includegraphics[scale = 0.60]{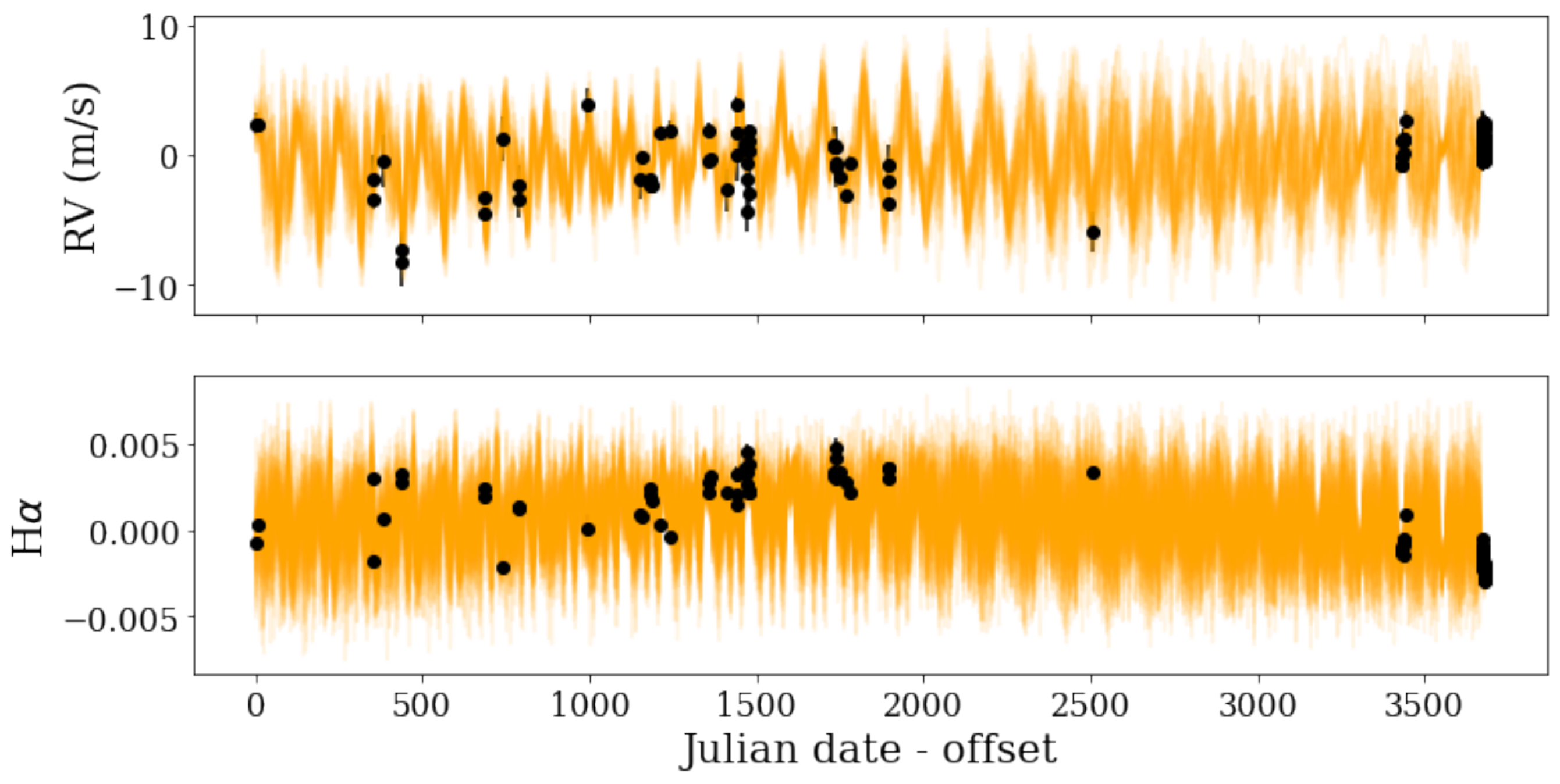}
    \end{tabular}
    \caption{{\bf Top two panels, blue:} Here we examine a single realization of the best-fit rotation model, plotted on top of the observed RV and H$\alpha$ data. The model includes harmonics of the true 125-day rotation period, which explains why \citet{escude14} found multiple peaks in the RV periodogram. We zoom in to times between 1100 and 1900 days after the first observation to show the behavior of the rotation model on shorter timescales. {\bf Bottom two panels, orange:} To illustrate the range of waveforms described by the best-fit GP hyperparameters, we plot 50 realizations of the best-fit rotation model over the duration of the time series.
    } 
    \label{fig:qsireal3}
    \end{center}
\end{figure}

To visualize the ensemble of time series described by the GP, we plot 50 randomly chosen pairs of RV and H$\alpha$ realizations of the best-fit rotation model (Figure \ref{fig:qsireal3}, orange). The figure also contains one randomly chosen realization of each model time series focused on a subset of the data (blue), illustrating the high-frequency components of the rotation signal. From the top of Figure \ref{fig:qsireal3}, we can trace both the periodic component and the evolution of the waveform over the course of the characteristic active-region lifetime $\lambda$. The existence of high-frequency components with periods $P < P_{\rm rot}$ explains why \citet{escude14} found more than one peak in the periodogram of the RVs.

\begin{figure}
    \centering
    \includegraphics[width=0.8\textwidth]{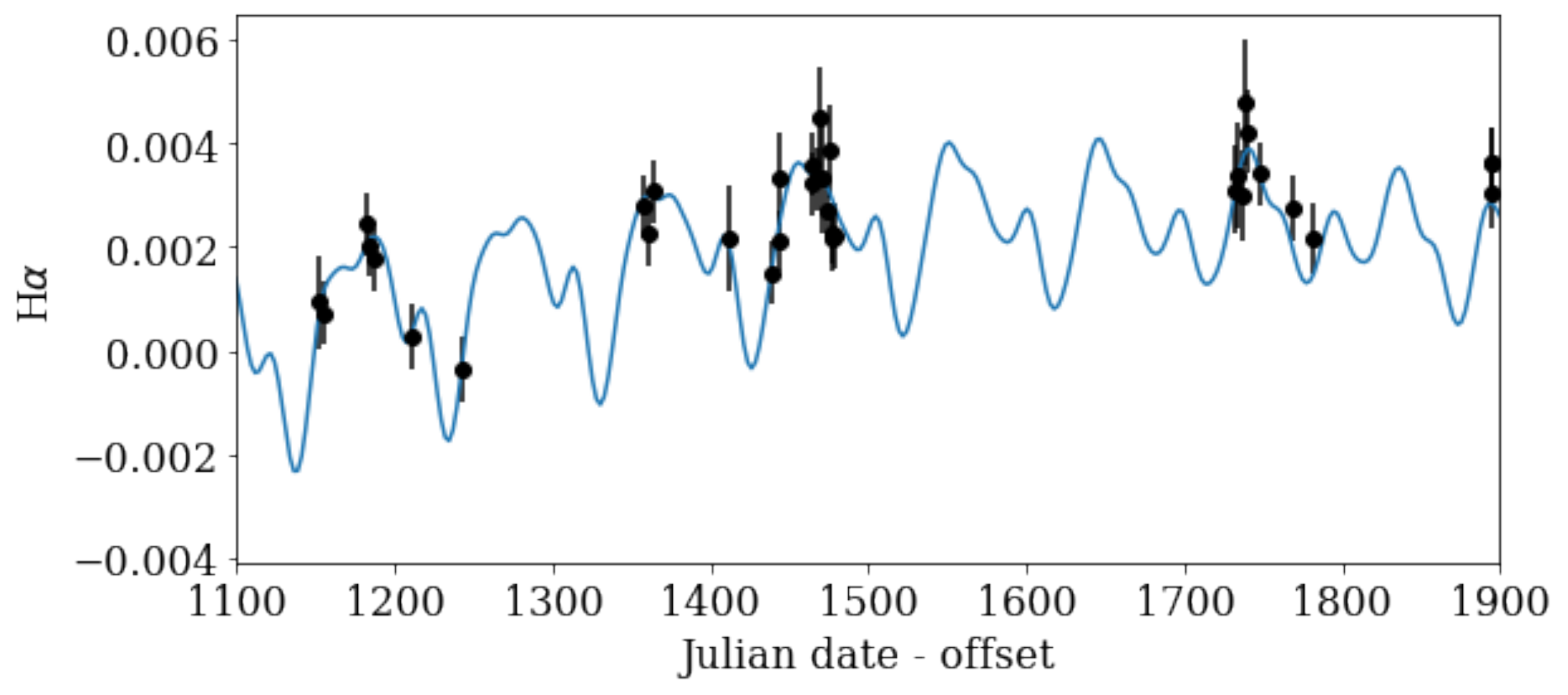}
    \caption{A single realization of the best-fit GP given the \citet{robertson15} H$\alpha$ dataset with error bars artificially inflated by a factor of 2.}
    \label{fig:ha_smooth}
\end{figure}

The high value of $\Gamma_{H\alpha}$ generates model H$\alpha$ time series that are much ``rougher'' (i.e.\ contain higher-frequency components) than the model RV time series. It is possible that the H$\alpha$ models are rough because the uncertainties in the H$\alpha$ data are underestimated.  \citet{robertson15} use the H$\alpha$ index definition of \citet{kurster03},
\begin{equation}
    I_{H\alpha} = \frac{F_{H\alpha}}{F_1 + F_2},
    \label{eq:haind}
\end{equation}
where $F_{H\alpha}$ is the flux in a band of width 1.6\AA\  centered at 6562.828\AA\ \citep{gomesdasilva11} and $F_1$ and $F_2$ are reference bands on either side of the H$\alpha$ line. 
$I_{H\alpha}$ uncertainties are calculated as in \citet{robertson13}, who include Poisson noise (see their Equation 2) but do not consider the noise introduced by interpolating the spectra to find the edges of the line core band and reference bands. If we artificially multiply the H$\alpha$ error bars by 2 (i.e.\ quadruple the variance), the best-fit value of $\Gamma_{H\alpha}$ drops to 3.2. Figure \ref{fig:ha_smooth} shows a sample realization of the resulting H$\alpha$ GP, which produces much smoother model time series.

\begin{figure} 
    \begin{center}
    \includegraphics[scale = 0.6]{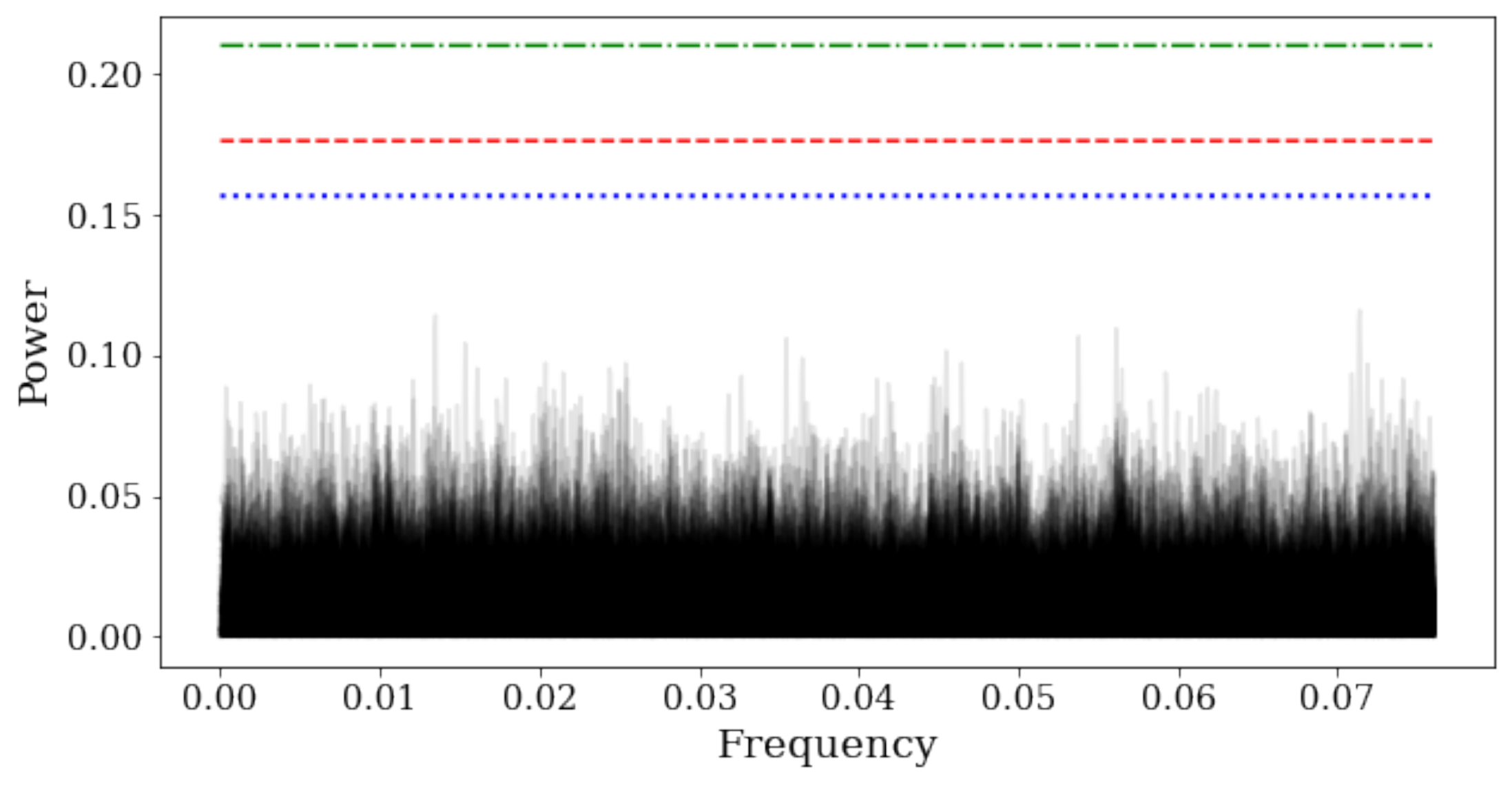}
    \caption{RV periodograms constructed from residuals of the quasi-periodic model. 
    The blue dotted, red dashed, and green dash-dot lines indicate 10\%, 5\%, and 1\% false alarm probabilities, respectively. We see no residual signals at the 121.5- and 48.6-day planet periods reported by \citet{escude14}.
    }
    \label{fig:qsipdgm}
    \end{center}
\end{figure}

To verify that the radial velocities of Kapteyn's star only provide evidence for rotation (\S \ref{subsec:noplanet}), we examined the residuals of the best-fit GP model. 
We randomly chose 300 RV realizations from the GP and subtracted them from the observational data to get 300 sets of residuals. Lomb-Scargle periodograms were created for each individual set of RV residuals and plotted together in Figure \ref{fig:qsipdgm}. The periodograms show no peaks with a low enough false-alarm probability to indicate an orbiting planet: all residual periodograms have FAP~$>$~10\% at all periods traced by the dataset. The strongest residual signal corresponds to a period of 14.17 days with a FAP of 35.9\%. In addition, the reported 48.6 and 121.5 day orbital periods \citep{escude14} and the 143-day rotational period found by \citet{robertson15} did not show any significant peaks in the residuals. We conclude that the only signal that is truly present in the data is the 125-day rotation period, and that the previously reported signals are spurious results of modeling quasi-periodic variation with truly periodic functions.

\section{Discussion and conclusions}\label{sec:disc}

It can be difficult to separate RV signals resulting from stellar activity and those resulting from orbiting planets. In the case of Kapteyn's star, there is argument over whether its reported planets are genuine or merely products of stellar activity. 
We used two GPs with joint, quasi-periodic kernels to simultaneously model H$\alpha$ and RV variations, with the goal of removing any RV signals caused by rotation and stellar activity.
The GP quasi-periodic model gives a rotational period for Kapteyn's star of 124.7 days and a correlation timescale, or characteristic active-region lifetime, of 694.1 days.
Any resulting significant signals that occur in the RV residuals would indicate the presence of exoplanets.
As we find no such signal, we conclude that Kapteyn's star does not host any exoplanets and that the signals of the reported planets were products of the star's rotation and activity.

However, a limitation of this work is the sparsity of data. Between the HARPS and HIRES data sets, there are only 112 data points spread over a 10 year period with a number of long pauses between observations. It is possible that there is simply not enough data for any periodic signal to be considered significant. As acknowledged by \citet{robertson15}, further and more regular observation of Kapteyn's star is needed to definitively determine the existence of the star's reported exoplanets. Moreover, previous work by \citet{feng16} and \citet{ribas18} suggests that GP regressions could interpret real signals as noise. These works indicate that, due to their flexibility, GPs may be prone to overfitting the noise, resulting in fewer and/or less significant signals in the residuals. Using a simpler kernel may help reduce the loss of significant signals from overfitting. However, this must be balanced with the need for flexibility to properly account for changing and shifting stellar activity cycles, which can result in false positives if not properly modeled.

There are similar works which also utilize GP for exoplanet searches. One such investigation of $\alpha$~Cen~B comes from \citet{rajpaul15}, who use GP to model the $\Delta$RV mean function with the equation $a + bt + ct^2$ where $a$, $b$, and $c$ are hyperparameters and the quadratic is designed to fit the binary motion. They model the covariance using the $F\dot{F}$ spot model, where $F$ is the spot coverage fraction of the visible hemisphere and $\dot{F}$ is its time derivative \citep{aigrain12}. Their tests of the GP model on datasets simulated by SOAP 2.0 \citep{dumusque14} were successful at distinguishing various activity cycles from planetary signals. In addition, they jointly model the  emission of calcium II H \& K lines and the cross correlation bisector inverse slope (BIS).

A key difference between our work and the work done by Rajpaul et al.\ is that our framework in which the RV and H$\alpha$ kernels share a period and decorrelation timescale, but are not otherwise constrained, allows for the H$\alpha$ and RV signals to phase-shift relative to one another. \citet{hatzes18} demonstrated that stellar noise can phase-shift: their observations of $\gamma$~Dra revealed a periodic signal that persisted for 8 years and was believed to be planetary in origin, until it disappeared for a few years before returning with a phase shift \citep[see also][]{ramirezdelgado20}. 
There are many ways to partially decouple the activity-indicator and RV signals---for example, a prominence on the star's limb would contribute much less to the velocity signal than to the H$\alpha$ signal.
Thus we prefer a model that can account for any drifting in the signal, or mismatching of phase in RV and H$\alpha$ data. While the $F \dot{F}$ model is physically motivated and insightful, it may be too deterministic to fully account for the stochastic appearance and disappearance of individual active regions.

Our approach also differs from that of \citet{haywood14}, who applied GP to the star CoRoT-7 to model two known planets after removing stellar activity from the signal. However, they operated under the assumption that the frequency structures in the RV and light curve data are the same. While the two are certainly similar, active regions in different locations on the star cause different RV variations, and there is enough uncertainty about exactly where and when an active region will appear that we prefer to allow a more flexible link between H$\alpha$ and RV.
The radial velocity variations caused by active regions are chromatic \citep{baroch20}, so we can't expect perfect coupling between RV and H$\alpha$ signals (i.e.\ a straight-line relationship between the two) if RVs are measured from the entire visible spectrum. \citet{morris20} experimented with detecting sunspots by cross-correlating a high-resolution stellar template with archival HARPS solar spectra. Their technique yielded some nondetections of spots even when the Sun was spotted, indicating that spectra are imperfect activity tracers. In a related study, \citet{cretignier20} found that in the 89\% of absorption lines that are to some degree sensitive to convective blueshift, the radial velocity shift induced by stellar activity is inversely proportional to the line depth. According to Cretignier et al., one can correct out the activity signature by mapping RV as a function of depth in the stellar atmosphere: planetary RV signals come primarily from the photosphere, while activity signals are primarily chromospheric. It is quite possible that the Kapteyn's star RV and H$\alpha$ time series are tracing different layers of the star's atmosphere.
By not imposing assumptions about the frequency structures of the H$\alpha$ and RV signals, we avoid eliminating valid possibilities for the relationship between the two signals.

Our work adds to the body of literature showing that GPs are effective in disentangling stellar activity signals from legitimate exoplanets, though concerns about overfitting the noise should be investigated further. 
As planet hunters dig deep into the stellar jitter,
more sophisticated techniques will be needed for screening new planet candidates.
We suggest using stars with both known rotation periods and confirmed exoplanets to test 
which GP kernel models are most reliable for planet validation, extending the work done by \citet{jones17} and \citet{pope16}. Planet searches would also benefit from further investigation of the relationship between RV and H$\alpha$ (and other activity indicators): we find that both are well modeled by the same underlying period, but their correlation is not strictly described by a straight-line model. 
Finally, planet hunters can continue to develop physically motivated GP kernels for other stellar noise sources besides rotation, such as magnetic activity cycles.

Acknowledgments: 
We thank Dan Foreman-Mackey for validating our likelihood function, Jessi Cisewski for input on Bayesian model fitting, and Alex Wise for discussions of the RV measurements and window function. Work by AB and HF was supported by a NASA EPSCOR RID seed grant to SDR. Work by VRD was supported by University of Delaware's Summer Scholars Program.


\end{document}